\DocumentMetadata{%
	pdfstandard=A-3u,
	pdfversion=1.7,
	lang=en-US,
}

\documentclass[balance,colorlinks,upint,subscriptcorrection,varvw,mathalfa=cal=boondoxo,german,spanish,vietnamese,russian,greek, nofoot]{asmeconf}

\hypersetup{%
	pdfauthor={John H. Lienhard},									  
	pdftitle={ASME Conference Paper LaTeX Template},                  
	pdfkeywords={ASME conference paper, LaTeX template, BibTeX style},
	pdfsubject = {Describes the asmeconf LaTeX template},			  
	pdflicenseurl={https://ctan.org/pkg/asmeconf},
}

\begin{document}


\ConfName{Proceedings of the ASME 2026\linebreak Fluids Engineering Division Summer Meeting}
\ConfAcronym{FEDSM2026}
\ConfDate{July 26-29, 2026} 
\ConfCity{Bellevue, WA} 
\PaperNo{FEDSM2026-184124}


\title{High-Order CFD Modeling of Rotating Actuator Line Systems} 
 
%
%
%

\SetAuthors{%
	Abdullah Al Imran\affil{1},  Meilin     Yu\affil{1}\CorrespondingAuthor{mlyu@umbc.edu}
	}
\SetAffiliation{1}{Department of Mechanical Engineering, University of Maryland Baltimore County (UMBC), Baltimore, MD, 21250, USA }



\maketitle





\begin{abstract}
High-order computational fluid dynamics (CFD) methods combined with actuator-based blade representations provide an attractive approach for simulating rotating energy systems. In this work, we develop a high-order computational framework that couples the flux reconstruction/correction procedure via reconstruction (FR/CPR) formulation with a rotating actuator line model (ALM). Blade rotation and unsteady aerodynamic forces are implemented through time-dependent source terms applied on a fixed cartesian grid. This approach allows simulations of a fully rotating turbine without using explicit blade geometry. Aerodynamic loads are computed using quasi-steady airfoil data and distributed into the flow field with Gaussian smoothing to maintain numerical stability in the high-order scheme. The framework is tested using a two-bladed vertical-axis wind turbine (VAWT) operating at a low tip speed ratio (TSR). The phase-averaged blade motion and overall load are analyzed to study variations with azimuthal angle. In addition, instantaneous and mean flow fields are examined to describe wake structure and unsteady vortex shedding. Normalized mean wake velocity profiles are used for validation through comparison with experimental measurements and blade-resolved large-eddy simulations (LES) reported in the literature. The results show that the proposed high-order virtual-body framework can reliably capture major blade-loading trends, flow organization, and wake features under strongly unsteady conditions, while maintaining geometric simplicity.
\end{abstract}





\section{Introduction}
Wind energy has become one of the most widely adopted renewable energy technologies because of its large-scale potential and continued cost reduction. Although horizontal-axis wind turbines (HAWTs) dominate commercial applications, VAWTs remain an active topic of research. This interest is driven by their ability to operate independently of wind direction, simpler mechanical design, and potential benefits in complex or space-limited environments~\cite{hand2021aerodynamic}. In recent years, VAWTs have also been proposed for tightly spaced counter-rotating array layouts, which may achieve higher power density than traditional wind farms~\cite{dabiri2011potential}. Despite these advantages, the aerodynamics of VAWTs are inherently complex. They involve strong changes in blade-relative velocity with rotation angle, dynamic stall, and highly unsteady interactions between blades and their wakes~\cite{simao2009visualization,scheurich2011effect}. Addressing these challenges requires numerical methods that can capture transient flow behavior while remaining efficient enough for parametric studies and simulations of turbine arrays.

Therefore, CFD has become an essential tool in wind-energy research, enabling systematic investigation of turbine aerodynamics and wake evolution. Reynolds-averaged Navier-Stokes (RANS) and unsteady RANS (URANS) approaches have been widely used for both HAWTs and VAWTs due to their relatively low computational cost, with careful attention paid to turbulence modeling, domain size, and temporal resolution~\cite{ferreira20072d,rezaeiha2017cfd}. LES has further advanced the field by resolving energetic turbulent structures and providing deeper insight into wake recovery, vortex shedding, and turbine--turbine interactions~\cite{stevens2017flow,porte2020wind}. Although these approaches have shaped much of the current understanding of wind-turbine aerodynamics, their predictive accuracy for vortex-dominated unsteady flows remains sensitive to numerical dissipation and modeling assumptions.

Actuator-based force modeling has emerged as a key enabling technology for wind turbine simulation by allowing turbine effects to be represented without body-fitted meshes. The actuator disk model (ADM) approximates the rotor as a distributed momentum sink and has been used extensively in studies of the wind-farm and atmospheric boundary-layer~\cite{calaf2010large,wu2011large}. ALM extends this concept by representing individual blades as rotating lifting lines, thereby capturing unsteady loading, azimuthal asymmetry, and near-wake vortex dynamics at moderate computational cost~\cite{sorensen2002numerical,mikkelsen2003actuator,troldborg2009actuator,churchfield2012large}. Comparative studies have shown that ALM significantly improves near-wake predictions relative to ADM, while remaining orders of magnitude cheaper than fully blade-resolved simulations~\cite{martinez2015large,stevens2018comparison}. A critical component of ALM is the regularization of blade forces through spatial smoothing kernels, which directly influences numerical stability and wake resolution.

High-order CFD methods, such as Flux Reconstruction (FR) \cite{huynh2007flux,wang2009unifying,vincent2011new}, Discontinuous Galerkin (DG) \cite{Cockburn_DG_1989,bassi1997high}, and Spectral Difference (SD) methods \cite{liu2006spectral} aim to overcome the limitations of these traditional low-order schemes by providing higher accuracy per degree of freedom. These high-order methods provide a better capture of complex flow phenomena while preserving computational efficiency. In the last decade, they have been applied to high-fidelity simulations of many types of vortex-dominated and turbulent flows, such as flapping wing/foil aerodynamics~\cite{PerssonEtAl_2012_IJNME,YEtAl_2013_JFS,YuEtAl_2018_JFS}, fluid-structure interaction~\cite{FroehlePersson_2014_JCP,LiuEtAl_2022_AMM}, aerodynamic load regulation in highly unsteady environments~\cite{poudel2021gust,zozimo2025spectral}. Several studies have demonstrated the potential of high-order methods for wind-turbine wake simulations, including LES-capable formulations that better preserve coherent structures in the near and intermediate wake regions~\cite{Frere2016DG_WT,Kleusberg2017SEM_Wakes}. Recently, high-order FR methods combined with LES have been applied to wind turbine flows and have shown improved resolution of aerodynamic performance and wake structures in complex turbine configurations~\cite{ding2023high}. These studies indicate that high-order formulations are effective in capturing vortex-dominated flow features while reducing numerical dissipation. At the same time, ALM has been integrated into high-order discontinuous Galerkin spectral element methods to improve the accuracy of wind turbine simulations~\cite{marino2024modelling}. Increasing the polynomial order in these frameworks has been shown to enhance wake resolution and turbulence representation without requiring finer meshes. Together, these advances suggest that high-order solvers are well suited to modeling the unsteady aerodynamics of VAWTs. And this is possible when turbine forcing and flow–structure coupling are introduced in a stable and consistent manner.

Building upon these developments, the present work represents a natural progression of a recently established source-term-based high-order computational framework originally developed for wind energy applications~\cite{imran2025high,imran2025development}. In these earlier studies, the authors developed and systematically validated a high-order FR/CPR solver coupled with actuator-based source-term representations on fixed cartesian grids. The methodology was verified using a set of standard test cases, including inviscid and viscous flow over fixed cylinders and viscous flow over rotating cylinders. The results demonstrate that properly regularized source terms can capture both steady and unsteady flow features for stationary and rotating bodies without requiring body-fitted meshes. Following these validations, the framework was applied to a fixed, steady VAWT using a non-rotating ADM. The results demonstrate that they accurately captured the mean momentum deficit and wake characteristics. However, ADM's are based on uniform or time-averaged forcing by design. As a result, they can not represent unsteady blade loading, wake asymmetry, or the detailed flow dynamics associated with rotating systems. The present study addresses this limitation by extending the high-order source-term formulation to fully rotating ALM, enabling time-resolved representation of blade motion and unsteady forcing within a fixed Eulerian framework. This extension establishes a general high-order methodology for rotating ALM systems, bridging previously validated static source-term models with time-accurate simulations of rotating actuators. In doing so, it provides a principled foundation for future investigations of complex multi-rotor and interacting wake configurations.

\section{Methodology}
\subsection{Flux Reconstruction Formulation}
\subsubsection{Governing Equations}
The two-dimensional (2D) unsteady compressible Navier-Stokes equations in the physical domain $(t,x,y)$ can be written as
\begin{equation}
\frac{\partial \mathbf{Q}}{\partial t}
+
\frac{\partial \mathbf{F}}{\partial x}
+
\frac{\partial \mathbf{G}}{\partial y}
=
0,
\label{eq:NS_conservative}
\end{equation}
where the vector of conservative variables is given by
\begin{equation}
\mathbf{Q}
=
\begin{pmatrix}
\rho \\
\rho u \\
\rho v \\
E
\end{pmatrix},
\end{equation}
with $\rho$ denoting the density of the fluid, $(u,v)$ the velocity components in the $x$ and $y$ directions, and $E$ the total energy per unit volume. For a perfect gas, the total energy is defined as
\begin{equation}
E = \frac{p}{\gamma - 1} + \frac{1}{2}\rho (u^2 + v^2),
\end{equation}
where $p$ is the pressure and $\gamma$ is the ratio of specific heats.

The inviscid and viscous flux vectors in the $x$ and $y$ directions are expressed as
\begin{equation}
\mathbf{F} = \mathbf{F}^{\mathrm{inv}} - \mathbf{F}^{\mathrm{vis}}, \qquad
\mathbf{G} = \mathbf{G}^{\mathrm{inv}} - \mathbf{G}^{\mathrm{vis}},
\end{equation}
with
\begin{equation}
\mathbf{F}^{\mathrm{inv}} =
\begin{pmatrix}
\rho u \\
\rho u^2 + p \\
\rho u v \\
u (E + p)
\end{pmatrix},
\qquad
\mathbf{G}^{\mathrm{inv}} =
\begin{pmatrix}
\rho v \\
\rho u v \\
\rho v^2 + p \\
v (E + p)
\end{pmatrix}.
\end{equation}

The viscous fluxes are given by
\begin{equation}
\mathbf{F}^{\mathrm{vis}} =
\begin{pmatrix}
0 \\
\tau_{xx} \\
\tau_{xy} \\
u \tau_{xx} + v \tau_{xy} + q_x
\end{pmatrix},
\qquad
\mathbf{G}^{\mathrm{vis}} =
\begin{pmatrix}
0 \\
\tau_{yx} \\
\tau_{yy} \\
u \tau_{yx} + v \tau_{yy} + q_y
\end{pmatrix},
\end{equation}
where $\boldsymbol{\tau}$ denotes the viscous stress tensor and $(q_x, q_y)$ are the heat flux components. For a Newtonian fluid, the stress components are defined as
\begin{align}
\tau_{xx} &= 2\mu \left( \frac{\partial u}{\partial x} - \frac{1}{3} \nabla \cdot \mathbf{u} \right), \\
\tau_{yy} &= 2\mu \left( \frac{\partial v}{\partial y} - \frac{1}{3} \nabla \cdot \mathbf{u} \right), \\
\tau_{xy} &= \tau_{yx} = \mu \left( \frac{\partial u}{\partial y} + \frac{\partial v}{\partial x} \right),
\end{align}
and the heat fluxes are computed using Fourier’s law,
\begin{equation}
q_x = -\frac{\mu C_p}{\mathrm{Pr}} \frac{\partial T}{\partial x}, \qquad
q_y = -\frac{\mu C_p}{\mathrm{Pr}} \frac{\partial T}{\partial y},
\end{equation}
where $\mu$ is the dynamic viscosity, $C_p$ is the specific heat at constant pressure, $\mathrm{Pr}$ is the Prandtl number and $T$ is the temperature.\\

\begin{figure}[t]
\centering\includegraphics[width=\linewidth, alt = {}]{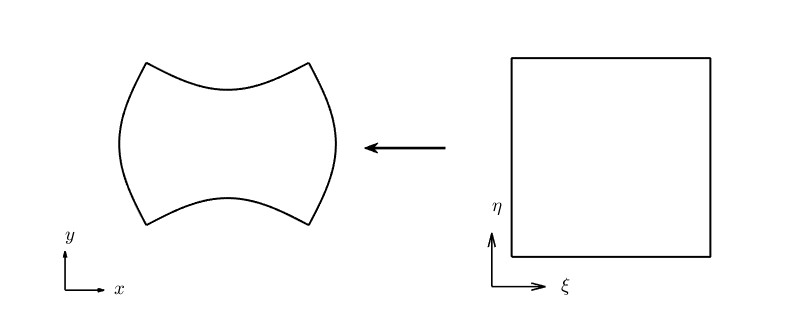}
\caption{Transformation from a physical domain to computational domain}
\label{fig:transformation}
\end{figure}

To apply FR/CPR discretization to a standard reference element, a time-dependent mapping is introduced from the physical domain $(t,x,y)$ to the computational domain $(\tau,\xi,\eta)$, where $\tau=t$ and
$(\xi,\eta)\in[-1,1]\times[-1,1]$, as illustrated in Fig.~\ref{fig:transformation}. 

Under this mapping, Eq.~\eqref{eq:NS_conservative} can be written in transformed form as
\begin{equation}
\frac{\partial \tilde{\mathbf{Q}}}{\partial \tau}
+
\frac{\partial \tilde{\mathbf{F}}}{\partial \xi}
+
\frac{\partial \tilde{\mathbf{G}}}{\partial \eta}
=
0,
\label{eq:NS_transformed}
\end{equation}
where $|J|$ is the Jacobian determinant of the transformation and
\begin{equation}
\tilde{\mathbf{Q}} = |J|\,\mathbf{Q}, \qquad
\end{equation}

The transformed fluxes $\tilde{\mathbf{F}}$ and $\tilde{\mathbf{G}}$ are defined by
\begin{align}
\tilde{\mathbf{F}} &= |J|\left(\mathbf{Q}\,\xi_t + \mathbf{F}\,\xi_x + \mathbf{G}\,\xi_y\right), \label{eq:F_flux} \\
\tilde{\mathbf{G}} &= |J|\left(\mathbf{Q}\,\eta_t + \mathbf{F}\,\eta_x + \mathbf{G}\,\eta_y\right). \label{eq:G_flux}
\end{align}

For the general time-dependent mapping $(x,y,t)=\mathbf{x}(\xi,\eta,\tau)$, the Jacobian matrix is defined as
\begin{equation}
\mathbf{J}
=
\frac{\partial(x,y,t)}{\partial(\xi,\eta,\tau)}
=
\begin{pmatrix}
x_\xi & x_\eta & x_\tau \\
y_\xi & y_\eta & y_\tau \\
0     & 0      & 1
\end{pmatrix},
\label{eq:jacobian}
\end{equation}
\begin{equation}
|J| = \det(\mathbf{J}) = x_\xi y_\eta - x_\eta y_\xi .
\end{equation}
and the inverse Jacobian is given by
\begin{equation}
\mathbf{J}^{-1}=
\frac{\partial(\xi,\eta,\tau)}{\partial(x,y,t)}
=
\begin{pmatrix}
\xi_x & \xi_y & \xi_t \\
\eta_x & \eta_y & \eta_t \\
0     & 0      & 1
\end{pmatrix}.
\label{eq:inv_jacobian}
\end{equation}
In this work, the grid is stationary (i.e., $x_\tau=y_\tau=0$), so $\xi_t=\eta_t=0$ and the transformed system reduces
to a purely spatial coordinate mapping while retaining the conservative form in Eq.~\eqref{eq:NS_transformed}.

\subsubsection{Spatial Discretization}
In the FR/CPR framework, the physical domain is partitioned into a set of non-overlapping elements, denoted by
$\Omega = \bigcup_{e=1}^{N_e} \Omega_e$, where $N_e$ is the total number of elements. Each physical element
$\Omega_e$ is mapped to a standard reference element $(\xi,\eta)\in[-1,1]\times[-1,1]$ using the coordinate
transformation introduced previously. Within each element, the solution is approximated by a polynomial of degree
$N$ in each spatial direction.

\begin{figure}[h]
\centering\includegraphics[width=\linewidth, alt = {}]{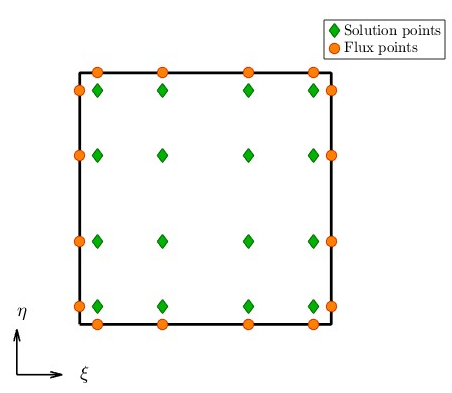}
\caption{Distribution of solution points (denoted by diamonds and) and flux points (denoted by circles) in a standard quadrilateral element for a third-order scheme}
\label{fig:solutionflux}
\end{figure}

The conservative solution vector $\tilde{\mathbf{Q}}(\xi,\eta,\tau)$ within an element is represented by an
element-local polynomial expansion,
\begin{equation}
\tilde{\mathbf{Q}}^h(\xi,\eta,\tau)
=
\sum_{i=1}^{N_s}
\tilde{\mathbf{Q}}_i(\tau)\,
\ell_i(\xi,\eta),
\end{equation}
where $\ell_i$ are Lagrange basis polynomials associated with a set of $N_s=(N+1)^2$ solution points.
In this work, the solution points are chosen as Gauss–Legendre points, which provide good quadrature accuracy and numerical stability, as shown in Fig.~\ref{fig:solutionflux}.

The corresponding fluxes are first evaluated locally at the solution points to construct the local flux polynomials,
\begin{align}
\tilde{\mathbf{F}}_{\mathrm{loc}}^h(\xi,\eta)
&=
\sum_{i=1}^{N_s}
\tilde{\mathbf{F}}(\tilde{\mathbf{Q}}_i)\,
\ell_i(\xi,\eta), \label{eq:Floc} \\
\tilde{\mathbf{G}}_{\mathrm{loc}}^h(\xi,\eta)
&=
\sum_{i=1}^{N_s}
\tilde{\mathbf{G}}(\tilde{\mathbf{Q}}_i)\,
\ell_i(\xi,\eta). \label{eq:Gloc}
\end{align}
where $\tilde{\mathbf{F}}$ and $\tilde{\mathbf{G}}$ denote the transformed fluxes defined in
Eqs.~\eqref{eq:F_flux}--\eqref{eq:G_flux}. These local flux polynomials are discontinuous across element interfaces
and therefore do not, by themselves, ensure conservation.

To enforce conservation and information exchange between neighboring elements, numerical fluxes are
introduced at element interfaces. The numerical fluxes are evaluated at a set of flux points located on the boundaries of the reference element. At each flux point, left and right states are reconstructed from the adjacent elements, and an approximate Riemann solver is used to compute the common inviscid flux. In this study, a Roe-type approximate Riemann solver is employed for inviscid fluxes~\cite{roe1981approximate},  while the common viscous fluxes at the cell interfaces are obtained using the approach developed by Bassi and Rebay~\cite{bassi1997high}. 
The discrepancy between the local fluxes and the common fluxes at element boundaries is then distributed back into
the element interior through correction functions. The corrected fluxes are written as

\begin{align}
\label{eq:corr1}
\tilde{\mathbf{F}}^h(\xi,\eta)
&=
\tilde{\mathbf{F}}_{\mathrm{loc}}^h(\xi,\eta)
+
\tilde{\mathbf{F}}_{\mathrm{cor}}^h(\xi,\eta), \\
\label{eq:corr2}
\tilde{\mathbf{G}}^h(\xi,\eta)
&=
\tilde{\mathbf{G}}_{\mathrm{loc}}^h(\xi,\eta)
+
\tilde{\mathbf{G}}_{\mathrm{cor}}^h(\xi,\eta),
\end{align}
where the correction fluxes $\tilde{\mathbf{F}}_{\mathrm{cor}}^h$ and
$\tilde{\mathbf{G}}_{\mathrm{cor}}^h$ are constructed such that the corrected flux polynomials exactly match the
numerical fluxes at the element boundaries. In this study, Radau polynomials, which can recover the DG scheme, are used as correction functions.

Upon substituting Eq.~\eqref{eq:corr1} and Eq.~\eqref{eq:corr2} into Eq.~\eqref{eq:NS_transformed}, the governing equations can be reformulated as:

\begin{equation}
\label{eq:semi-discrete}
\frac{\partial\tilde{\mathbf{Q}}}{\partial\tau}+(\frac{\partial\tilde{\mathbf{F}}_{\mathrm{loc}}^h}{\partial\xi}+\frac{\partial\tilde{\mathbf{G}}_{\mathrm{loc}}^h}{\partial\eta})+(\frac{\partial\tilde{\mathbf{F}}_{\mathrm{cor}}^h}{\partial\xi}+\frac{\partial\tilde{\mathbf{G}}_{\mathrm{cor}}^h}{\partial\eta})=0,
\end{equation}

This semi-discrete system Eq.~\eqref{eq:semi-discrete} is advanced in time using an explicit time-integration scheme, as described in the following Section~\ref{sec:temporal_discretization}.

\subsubsection{Temporal Discretization}
\label{sec:temporal_discretization}
Equation~\eqref{eq:NS_conservative} can also be written in the following form:
\begin{equation}
\label{eq:11}
\frac{\partial \mathbf{Q}}{\partial t}=\mathbf{R}(\mathbf{Q}, \nabla \mathbf{Q}),
\end{equation}
where $R$ stands for the residual. The time integration of Eq.~\eqref{eq:11} is performed using the explicit Strong Stability Preserving Runge-Kutta (SSP-RK) method \cite{shu1988total}. This method is designed to preserve the stability properties of the underlying spatial discretization, particularly for solving hyperbolic partial differential equations. The three-stage SSP-RK method is commonly defined by the following steps:

\begin{equation}
\label{eq:12}
\begin{cases}\mathbf{Q}^{(1)}=\mathbf{Q}^{n}+\Delta t \mathbf{R}^n\\ 
\mathbf{Q}^{(2)}=\frac{3}{4}\mathbf{Q}^{n}+\frac{1}{4}\mathbf{Q}^{(1)}+\frac{1}{4}\Delta t \mathbf{R}^{(1)}\\ 
\mathbf{Q}^{n+1}=\frac{1}{3}\mathbf{Q}^{n}+\frac{2}{3}\mathbf{Q}^{(2)}+\frac{2}{3}\Delta t \mathbf{R}^{(2)}\end{cases}
\end{equation}

This method ensures strong stability properties and high-order accuracy, making it suitable for advancing the solution in time while maintaining the stability of the spatial discretization. More details of the numerical framework can be found in our previous work~\cite{ywl14,wy20,wgy20}.

\subsection{Actuator Line Model for Vertical-Axis Wind Turbine}
The ALM method extends the classical theory of blade elements by representing turbine blades as rotating lifting lines embedded within the flow field. In our current high-order FR/CPR model, the aerodynamic forces induced by the turbine blades to the wind turbulence are computed using the ALM~\cite{sorensen2002numerical}.

\begin{figure}[h]
\centering
\includegraphics[width=\linewidth]{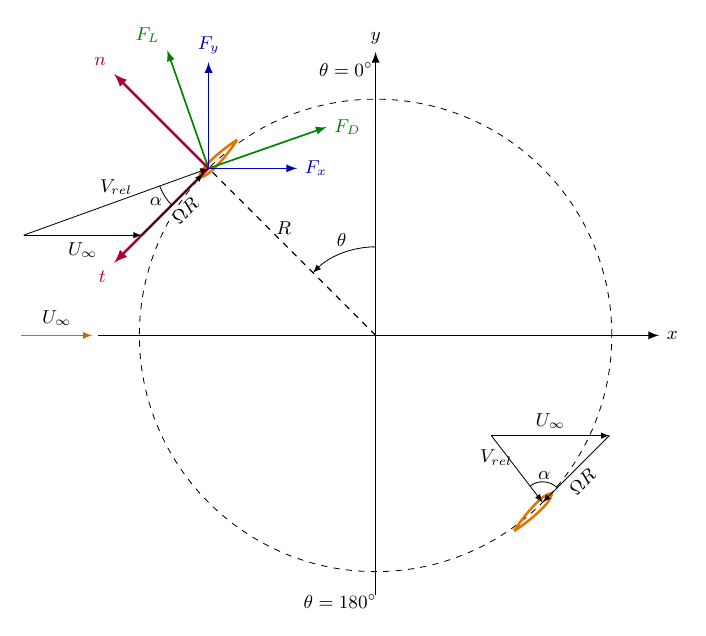}
\caption{Schematic of VAWT blade kinematics showing the azimuthal position, incoming flow, relative velocity, and the decomposition of aerodynamic forces in the local blade coordinate system defined by the tangential and radial directions.}
\label{fig:vawt_kinematics}
\end{figure}

We consider a horizontal cross-section of a blade rotating with angular velocity $\Omega$ at a fixed radial distance $R$ from the rotor center as shown on Fig.~\ref{fig:vawt_kinematics}. The incoming flow is represented by a uniform freestream velocity $\mathbf{U}_\infty$. We define the azimuthal position of the blade using the angle $\theta$, measured counter-clockwise. A local coordinate system is defined at each actuator point using a tangential direction $\hat{\mathbf{t}}$ and a radial direction $\hat{\mathbf{n}}$ pointing outward from the rotor center.

The relative velocity experienced by the blade is obtained from the combination of the freestream velocity and the blade rotational motion,
\begin{equation}
\mathbf{V}_{\mathrm{rel}} = \mathbf{U}_\infty - \Omega R \, \hat{\mathbf{t}},
\end{equation}
where $\hat{\mathbf{t}}$ is the unit vector in the tangential direction.

The angle of attack $\alpha$ is computed from the components of the relative velocity as
\begin{equation}
\alpha = \tan^{-1}\left(\frac{V_n}{V_t}\right),
\end{equation}
where $V_t$ and $V_n$ are the tangential and normal components of the relative velocity with respect to the blade chord.

The aerodynamic forces acting on the blade are expressed in terms of lift and drag components aligned with the relative velocity. Using quasi-steady airfoil data from~\cite{sheldahl1981aerodynamic}, the forces are computed as

\begin{equation}
F_L = \tfrac{1}{2}\rho |\mathbf{V}_{\mathrm{rel}}|^2 C \, C_l(\alpha),
\end{equation}
\begin{equation}
F_D = \tfrac{1}{2}\rho |\mathbf{V}_{\mathrm{rel}}|^2 C \, C_d(\alpha),
\end{equation}
where $C$ is the blade chord length. 

Lift $F_L$ and drag $F_D$ are first projected onto the blade-attached coordinate system $(\hat{\mathbf n}, \hat{\mathbf t})$ to obtain the normal and tangential components. These components are then transformed into the global cartesian frame  to form the blade force vector $\mathbf{F}_b = (F_x, F_y)$, which is used to construct the distributed body forces applied to the flow field. In the present study, dynamic stall effects are not included and the aerodynamic loads are determined solely from the instantaneous angle of attack. It should be noted that the present ALM formulation is implemented in a two-dimensional framework, where each blade is represented by a single actuator point corresponding to its cross-section. This differs from conventional three-dimensional ALM approaches, where blades are discretized into multiple spanwise elements. The present formulation can be extended to three-dimensional configurations in a straightforward manner.

\subsection{Coupling of ALM with FR/CPR}
In the present work, the effect of the ALM model is incorporated into the governing equations through additional source terms on the RHS of the 2D compressible Navier-Stokes equations. Equation~\eqref{eq:NS_conservative} can be rewritten in the following form:
\begin{equation}
\label{eq:GE-source}
\frac{\partial \mathbf{Q}}{\partial t}
+
\frac{\partial \mathbf{F}}{\partial x}
+
\frac{\partial \mathbf{G}}{\partial y}
=
\begin{bmatrix}
0 \\
S_x \\
S_y \\
S_E
\end{bmatrix}.
\end{equation}
where $S_x$, $S_y$, and $S_E$ denote the momentum and energy source terms associated with the ALM forcing.

After applying the FR/CPR spatial discretization to Eq.~\eqref{eq:GE-source}, the modified semi-discrete form will be as follows:
\begin{equation}
\frac{d\mathbf{Q}}{dt} = \mathbf{R} + \mathbf{S},
\end{equation}
where $\mathbf{R}$ denotes the spatial residual obtained from the FR/CPR discretization and $\mathbf{S}$ represents the discrete counterpart of the source terms $(0,S_x,S_y,S_E)^T$ in Eq.~\eqref{eq:GE-source}.

The formulation of the source terms used to couple ALM with our in-house FR/CPR solver is described below. To avoid the application of a singular force at discrete actuator locations, the sectional blade forces $\mathbf{F}_b$, defined in the previous subsection, are distributed onto the Eulerian grid using a Gaussian regularization kernel. For a blade located at $\mathbf{x}_b$, the force density at a spatial location $\mathbf{x}$ is defined as

\begin{equation}
\mathbf{f}(\mathbf{x}) = -\sum_{b=1}^{N_b} \mathbf{F}_b
\frac{1}{2\pi \epsilon^2}
\exp\!\left(-\frac{|\mathbf{x}-\mathbf{x}_b|^2}{2\epsilon^2}\right),
\end{equation}
where the negative sign enforces action--reaction consistency, such that the force applied to the fluid is equal and opposite to the force acting on the blade. The regularization width $\epsilon$ controls the spatial spread of the force and is selected proportional to the local grid spacing. This ensures a smooth projection of the force across high-order elements.

Within each FR/CPR element, the source term contribution is evaluated at the solution points using the Gauss-Legendre quadrature. For each quadrature point, the distance to each blade is computed, and the Gaussian kernel is applied within a finite cutoff radius. The resulting momentum source terms are

\begin{equation}
S_x = f_x, \qquad S_y = f_y,
\end{equation}
while the corresponding energy source term is computed consistently as the mechanical work done by the force,

\begin{equation}
S_E = \mathbf{f}\cdot\mathbf{u}.
\end{equation}
This formulation enables the incorporation of ALM forcing within the FR/CPR solver for rotating VAWT simulations and extends previously validated static source-term representations to fully dynamic ALM~\cite{imran2025development}.
\begin{figure}[t]
    \centering
    \begin{minipage}{0.9\linewidth}
        \centering
        \includegraphics[width=\linewidth]{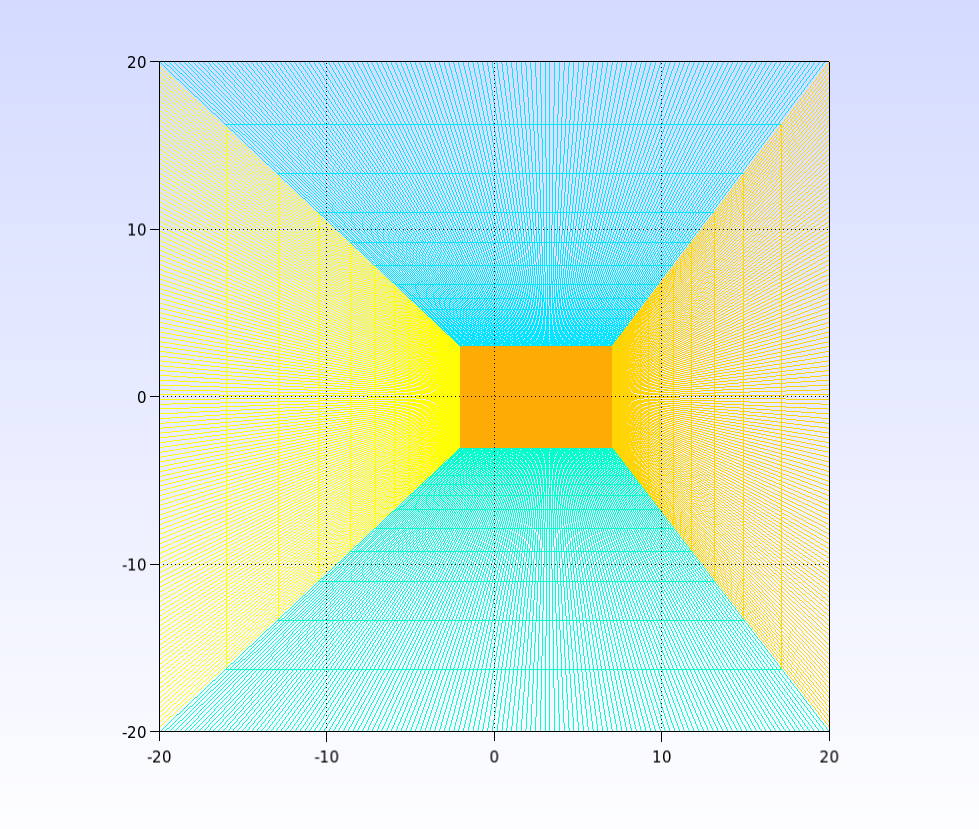}
        \\ (a) Full computational domain
    \end{minipage}
    \hfill
    \begin{minipage}{0.65\linewidth}
        \centering
        \includegraphics[width=\linewidth]{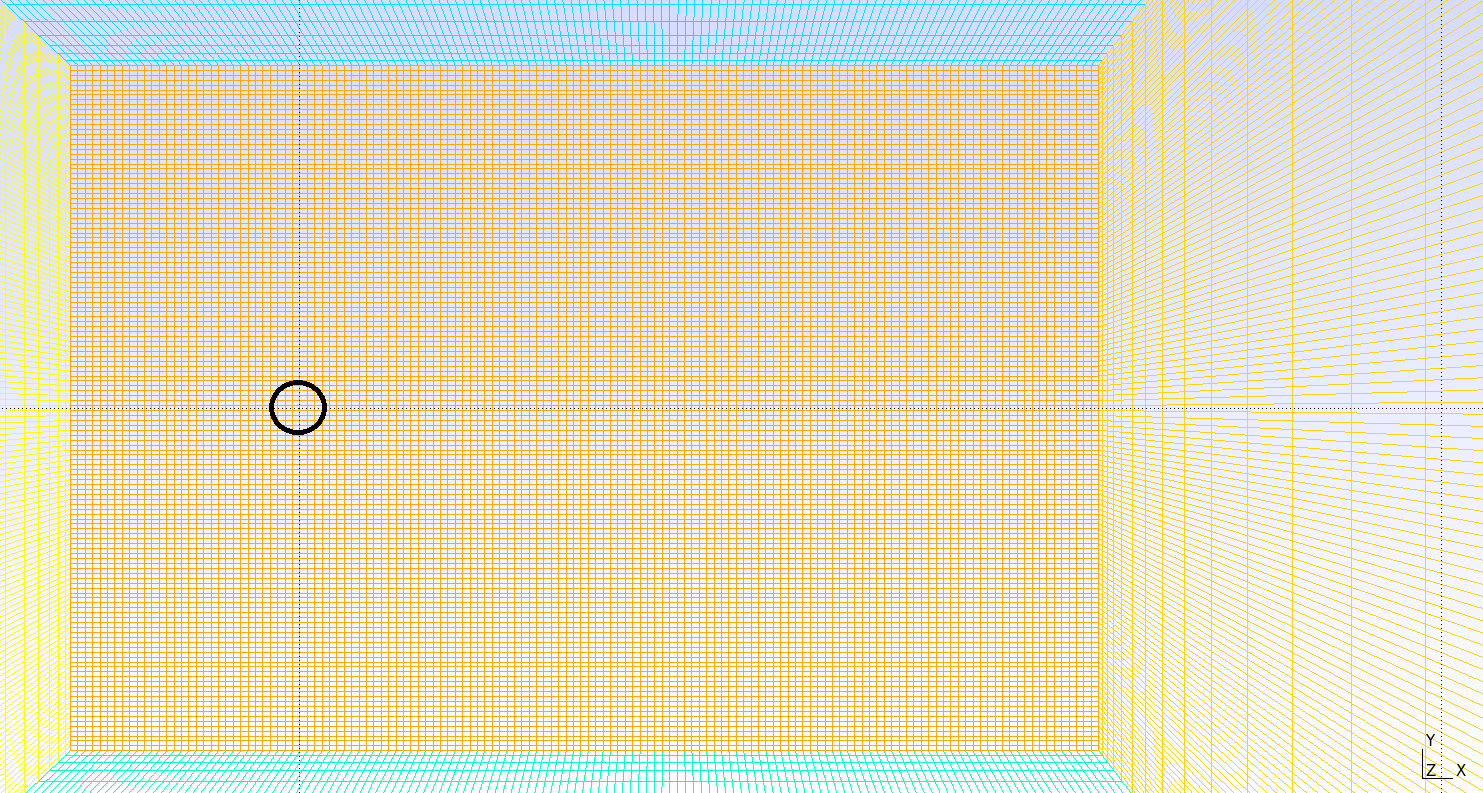}
        \\ (b) Zoomed view of the turbine-center region
    \end{minipage}
    \caption{Computational mesh of the domain with local refinement in the turbine-center region. The black circle with 0.5m radius indicates the turbine location and is shown for visualization only; no physical geometry is present in the mesh.}
    \label{fig:mesh}
\end{figure}

\subsection{Computational Setup}
\label{sec:computational_setup}
In this study, we consider two different grid resolutions to assess the sensitivity of the solution to mesh refinement. All simulations are carried out on a fixed, structured mesh composed of quadrilateral elements, where the effects of the rotating VAWT are introduced exclusively through volumetric ALM source terms. The computational domain is a two-dimensional rectangular region that extends over $x,y \in [-20D,\,20D]$, where $D$ denotes the diameter of the rotor. The actuator-line rotor is located in the center of the domain. A uniform freestream velocity $U_\infty$ is prescribed at the inlet, while a zero-gradient condition is applied at the outlet. The upper and lower boundaries are treated as slip walls. The domain size is chosen to be large enough to avoid any significant influence of the boundary conditions on the wake development.

To better resolve the actuator forcing and near-wake dynamics, we defined a locally refined rectangular region around the rotor and its immediate wake, spanning $x \in [-2D,\,7D]$ and $y \in [-3D,\,3D]$. Outside this region, the mesh is gradually coarsened toward the outer boundaries. The two grid resolutions correspond to inner-region discretizations of $139\times139$ and $199\times199$, resulting in approximately 29,885 and 62,685 total cells in the domain, respectively. A representative view of the  coarse mesh and the refined region is shown in Fig.~\ref{fig:mesh}.

The turbine considered here has rotor radius $R$ and blade chord length $C$, and rotates at a constant angular velocity corresponding to the prescribed tip-speed ratio. Blade forces are computed using static airfoil data for an NACA~0021 airfoil. The actuator-line forces are distributed onto the Eulerian grid using a Gaussian regularization kernel, with a width proportional to the local grid spacing. Each simulation progresses over time until a statistically periodic state is reached. Instantaneous flow fields are used to examine the unsteady vortex dynamics, while time-averaged quantities are obtained over a statistically steady interval and used for wake analysis and validation.

\section{Results}
The primary objective of the results presented in this section is to assess the ability of the proposed high-order virtual-body ALM framework to simulate fully rotating and unsteady turbine operation on a fixed mesh. The results are organized into three subsections: model validation, blade kinematics and loading, and flow-field structure with wake analysis. Quantitative validation is first performed against experimental near-wake measurements available in the literature, followed by an examination of blade loading characteristics and flow-field behavior predicted by the present framework.

\subsection{Model Validation}
For validation, we selected two reference datasets consisting of an experiment and a blade-resolved numerical simulation. The experimental measurements come from the towing-tank study of Bachant and Wosnik~\cite{bachant2013performance}, which reported detailed near-wake velocity profiles for a three-bladed VAWT operating at a tip-speed ratio of $\lambda = 1.9$. In addition, we use results from a blade-resolved LES simulation reported by Hezaveh et al.~\cite{hezaveh2017simulation} as a high-fidelity numerical reference. Although the main analysis in this work focuses on $\mathrm{TSR}=2$, we performed an additional simulation in $\lambda = 1.9$ to allow direct comparison with these benchmark datasets. We extract the normalized mean streamwise velocity profile $\overline{U}/U_\infty$ along the cross-stream direction at the downstream location $x/D = 1$.

\begin{figure}[t]
\centering
\includegraphics[width=\linewidth]{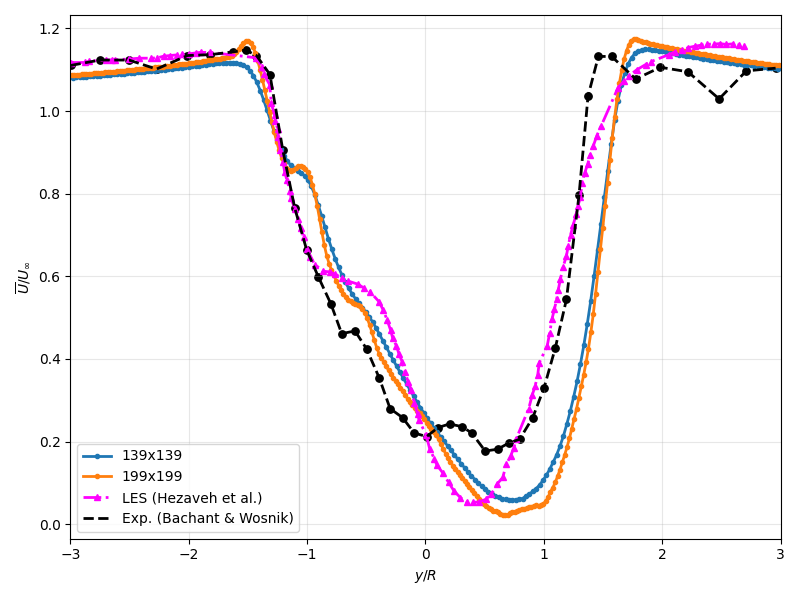}
\caption{Comparison of mean streamwise velocity cross-sections for $\lambda=1.9$ obtained using two grid resolutions together with experimental measurements~\cite{bachant2013performance} and blade-resolved LES simulation~\cite{hezaveh2017simulation} at a downstream distance of $2R$ from the turbine.}
\label{fig:grid_sensitivity}
\end{figure}

Figure~\ref{fig:grid_sensitivity} compares the velocity profiles obtained using the two grid resolutions with the experimental measurements and the blade-resolved LES results. The simulations capture the main features of the wake, including the velocity deficit and its lateral spread. As the grid is refined, the wake profile becomes slightly sharper in the core region while preserving the overall wake width. A grid convergence assessment shows that the solution becomes only weakly sensitive to further refinement, with minor differences observed between the two finest grids, indicating that the essential wake characteristics are well resolved.

\subsection{Blade Kinematics and Loading}
\begin{figure}[h]
\centering
\includegraphics[width=\linewidth]{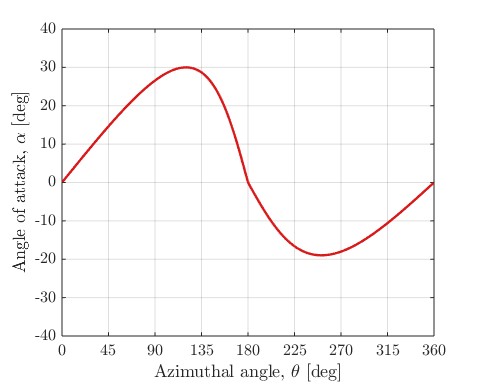}
\caption{Phase-averaged angle of attack as a function of azimuthal angle over one blade revolution.}
\label{fig:blade_analysis}
\end{figure}

Figure~\ref{fig:blade_analysis} shows the phase-averaged variation of the blade angle of attack as a function of the azimuthal position $\theta$ over one rotor revolution. The azimuthal angle is defined such that $\theta = 0^\circ$ corresponds to the blade located at the top of the rotor, with increasing $\theta$ indicating counter-clockwise rotation. The predicted angle-of-attack distribution exhibits a pronounced asymmetry between the upwind and downwind halves of the rotation. Large positive angles of attack occur in the upwind region ($0^\circ \lesssim \theta \lesssim 180^\circ$), while negative values appear during the downwind passage. This asymmetric behavior is a characteristic feature of low tip-speed-ratio operation, where strong variations in the relative velocity lead to highly unsteady blade aerodynamics.

\begin{figure}[!]
\centering
\includegraphics[width=\linewidth]{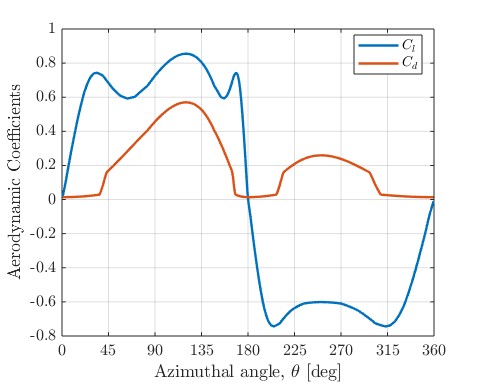}
\caption{Phase-averaged lift and drag coefficients as functions of the azimuthal angle.}
\label{fig:clcd}
\end{figure}

The corresponding phase-averaged lift and drag coefficients are presented in Fig.~\ref{fig:clcd}. Both coefficients follow the variation in angle of attack and exhibit strong azimuthal modulation. Elevated drag levels and reduced lift occur during the downwind passage, indicating the presence of separated-flow conditions associated with dynamic-stall-dominated operation. Although dynamic stall hysteresis is not explicitly modeled due to the use of static airfoil polars, the predicted trends reflect the expected aerodynamic behavior of vertical-axis turbines operating at low TSRs.

\begin{figure}[t]
\centering
\includegraphics[width=\linewidth]{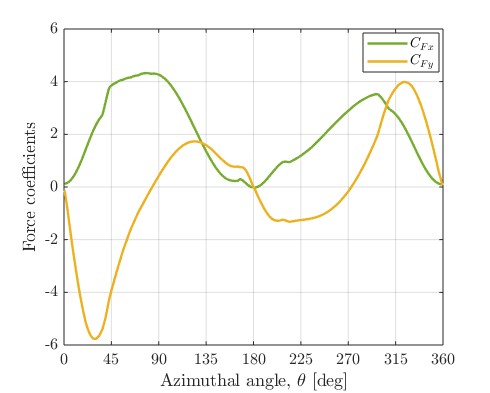}
\caption{Phase-averaged global force coefficients $C_{Fx}$ and $C_{Fy}$ as functions of the azimuthal angle.}
\label{fig:globalforcecoeff}
\end{figure}

The resulting global force coefficients acting on the blade are shown in Fig.~\ref{fig:globalforcecoeff}. Both $C_{Fx}$ and $C_{Fy}$ vary strongly with azimuthal position, with the largest force magnitudes occurring in the upwind portion of the rotation and reduced loading during the downwind passage. This phase-dependent loading reflects the combined influence of the asymmetric angle-of-attack distribution and the associated lift and drag variations.
\begin{figure*}[!]
\begin{subfigure}[b]{0.49\linewidth}
  \centering
  \includegraphics[width=\linewidth]{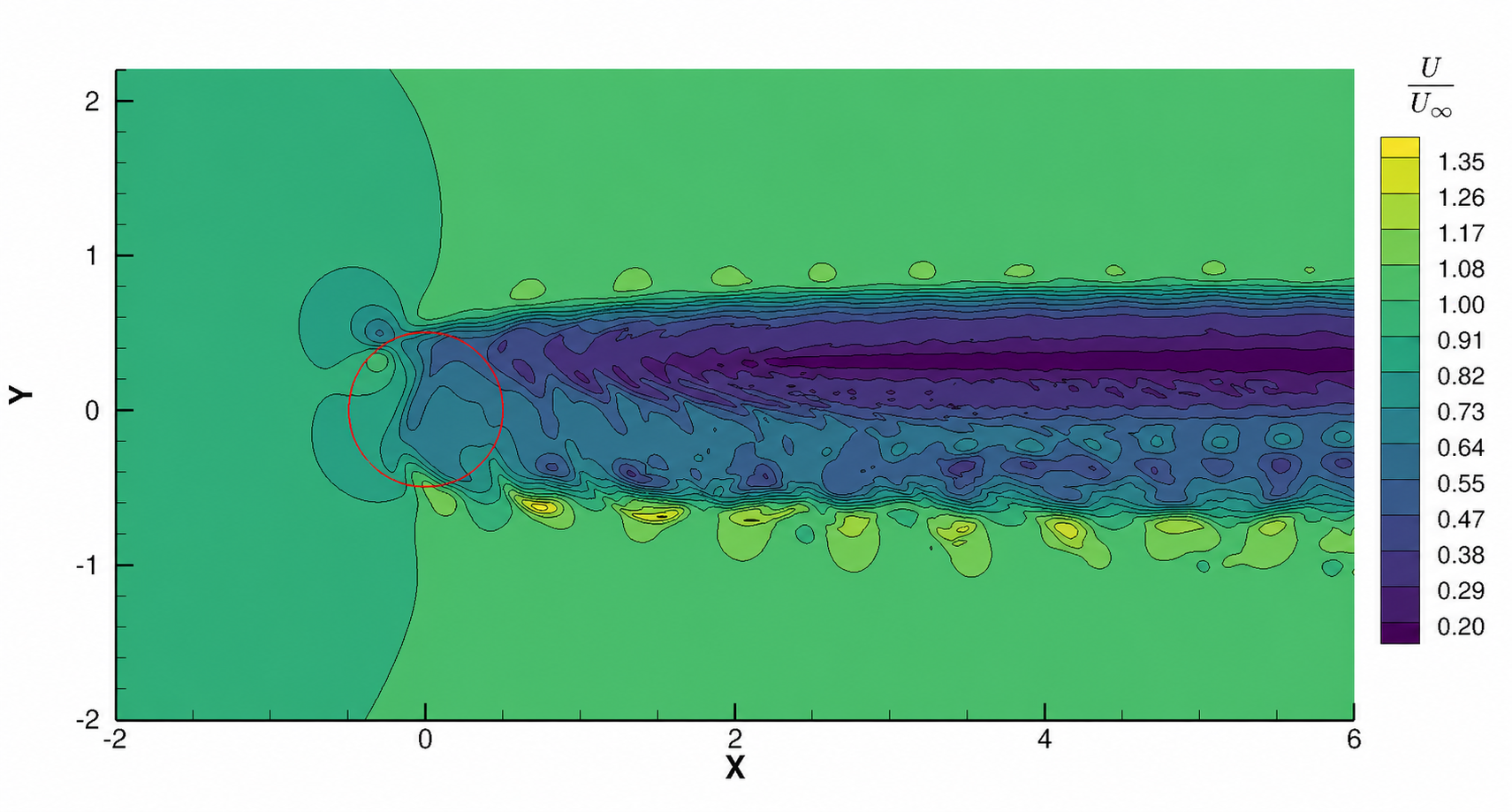}
  \caption{Instantaneous normalized streamwise velocity}
  \label{fig:inst_u}
\end{subfigure}\hfill
\begin{subfigure}[b]{0.50\linewidth}
  \centering
  \includegraphics[width=\linewidth]{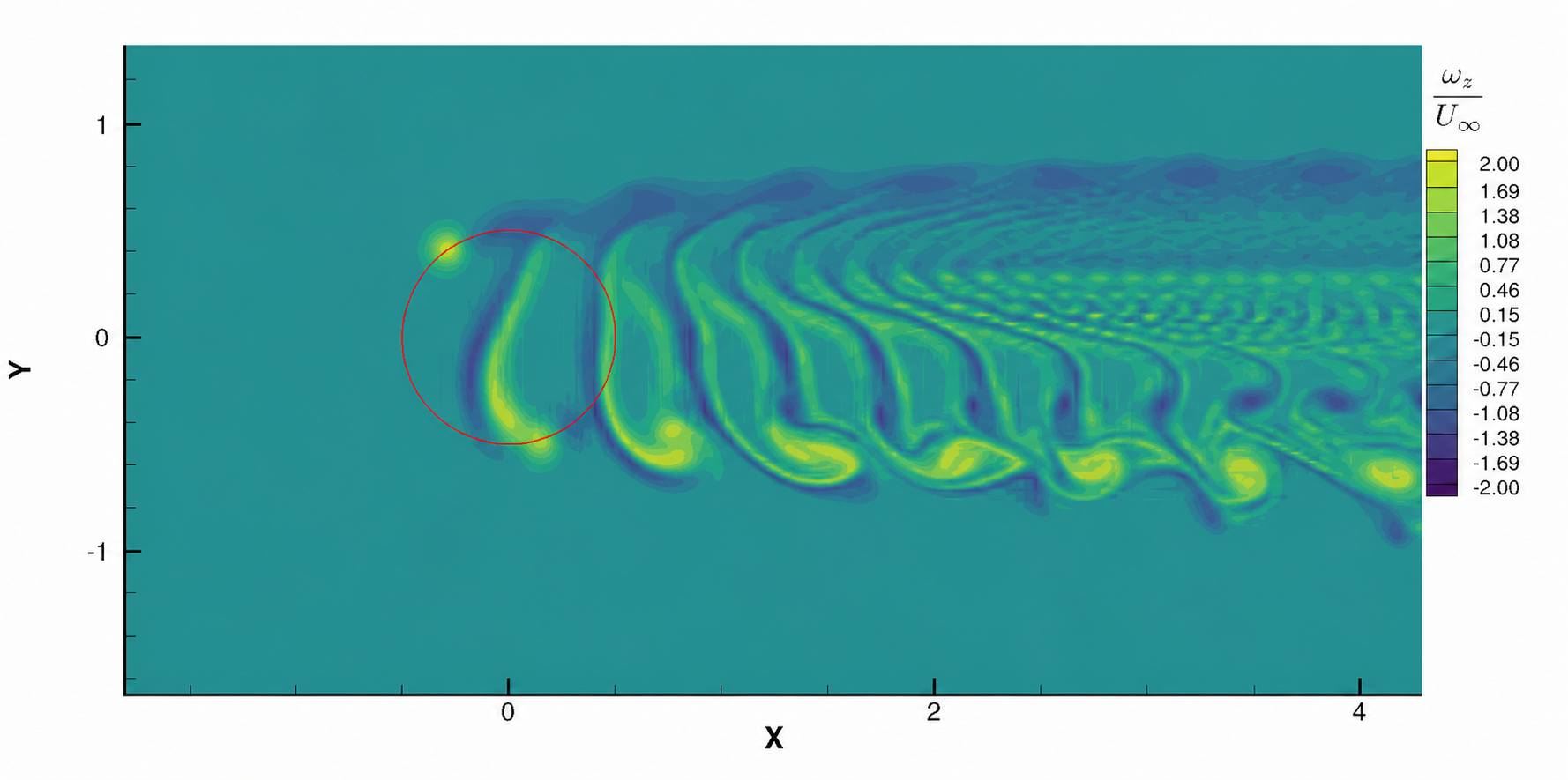}
  \caption{Instantaneous normalized spanwise vorticity}
  \label{fig:inst_vortz}
\end{subfigure}

\vspace{0.5em}

\begin{subfigure}[b]{0.49\linewidth}
  \centering
  \includegraphics[width=\linewidth]{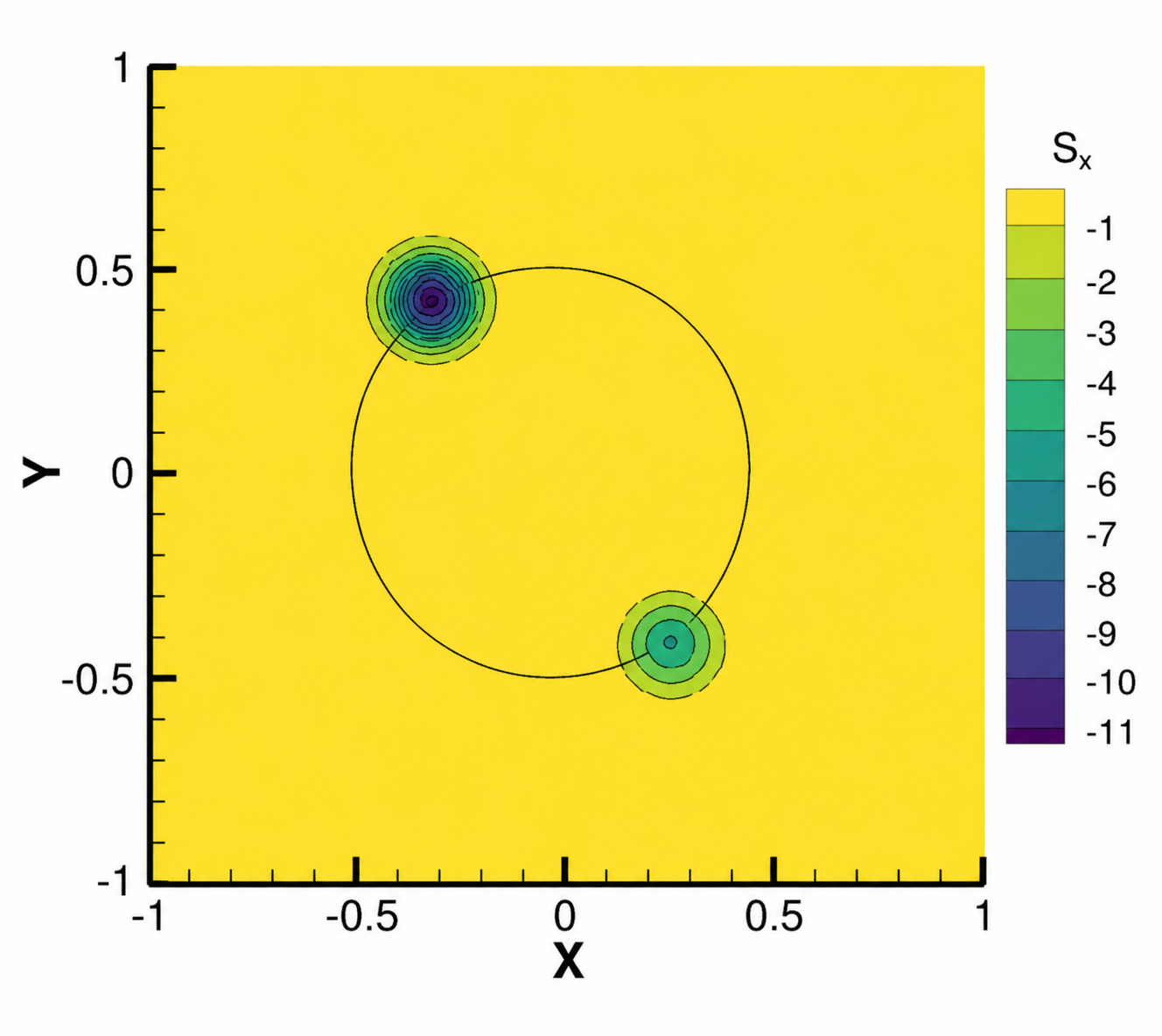}
  \caption{Streamwise momentum source term, $S_x$}
  \label{fig:sx}
\end{subfigure}\hfill
\begin{subfigure}[b]{0.48\linewidth}
  \centering
  \includegraphics[width=\linewidth]{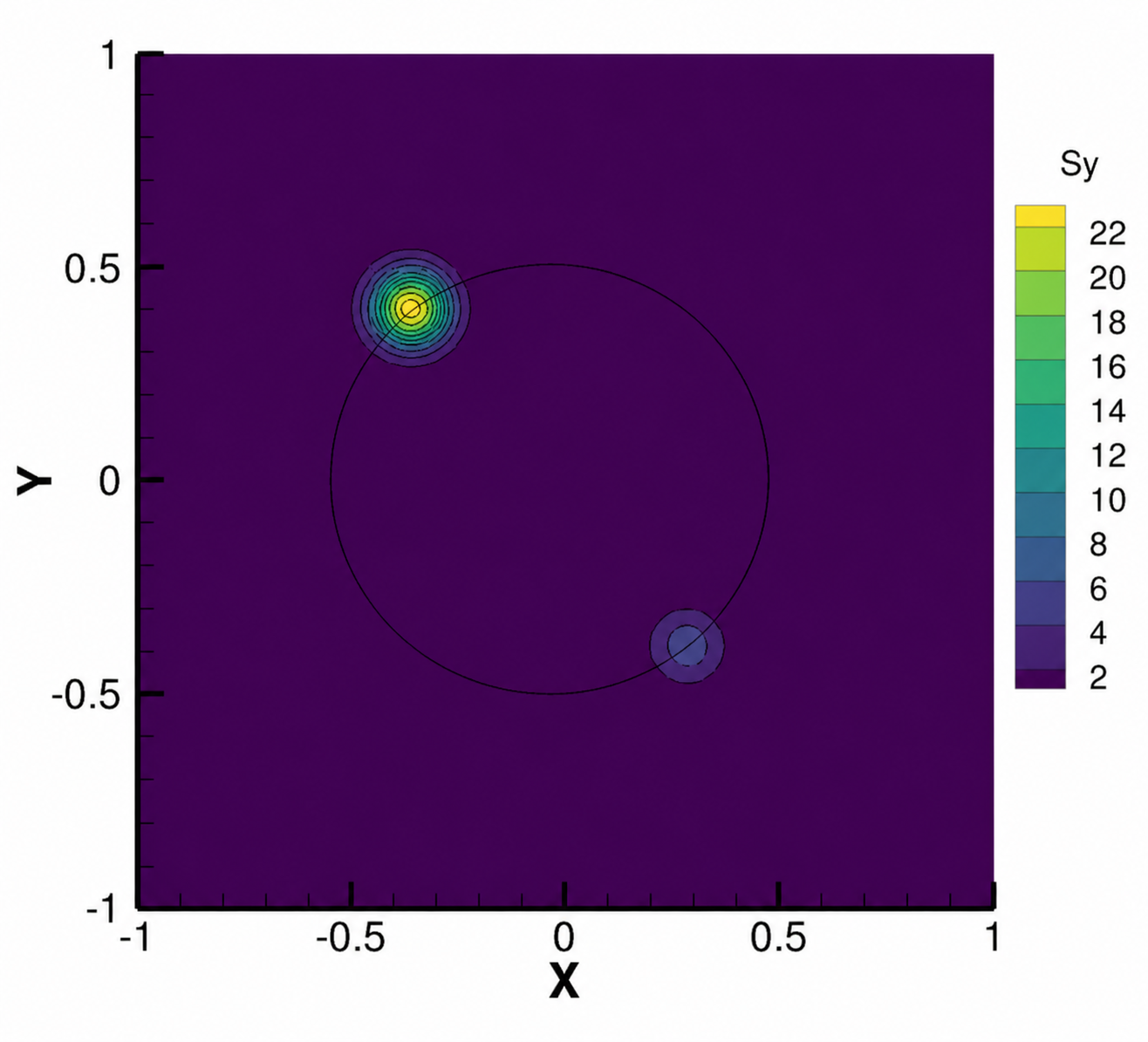}
  \caption{Cross-stream momentum source term, $S_y$}
  \label{fig:sy}
\end{subfigure}
\caption{Instantaneous flow and actuator-line forcing fields at $t=30$ s for the two-bladed turbine at $\mathrm{TSR}=2$. The top row shows the instantaneous flow response, while the bottom row shows the corresponding momentum source terms used to represent the blade forces. The circular outline indicates the projected turbine swept area $D=1$.}
\label{fig:inst_all}
\end{figure*}

\subsection{Instantaneous and Mean Flowfield Structure}
In this section, we apply the FR/CPR model coupled with ALM to simulate the flow induced by a two-bladed vertical-axis wind turbine. The turbine has a rotor radius $R=0.5$ m 
and a blade chord length $C=0.14$ m. The blades rotate in a counter-clockwise direction at a fixed tip-speed ratio $\lambda = \Omega R / U_\infty = 2.0$, which gives an angular velocity $\Omega = 4$ rad/s for the freestream velocity, $U_\infty = 1$ m/s.

The instantaneous flow field is examined at $t=30$ s, where the blades are located at $\theta_1 = 35.4^\circ$ and $\theta_2 = 215.4^\circ$. These azimuthal positions are reported modulo $360^\circ$, accounting for the periodic rotation of the blades over multiple revolutions. The simulation is performed with a time step $\Delta t = 10^{-4}$ using the coarse mesh with inner-region refinement, as described in Subsection~\ref{sec:computational_setup}. The ALM forcing is distributed using a Gaussian kernel with width $\epsilon = 2\Delta x$, where $\Delta x$ denotes the local grid spacing in the refined region.

\begin{figure*}[!]
  \centering
  \begin{subfigure}[b]{0.49\linewidth}
    \centering
    \includegraphics[width=\linewidth]{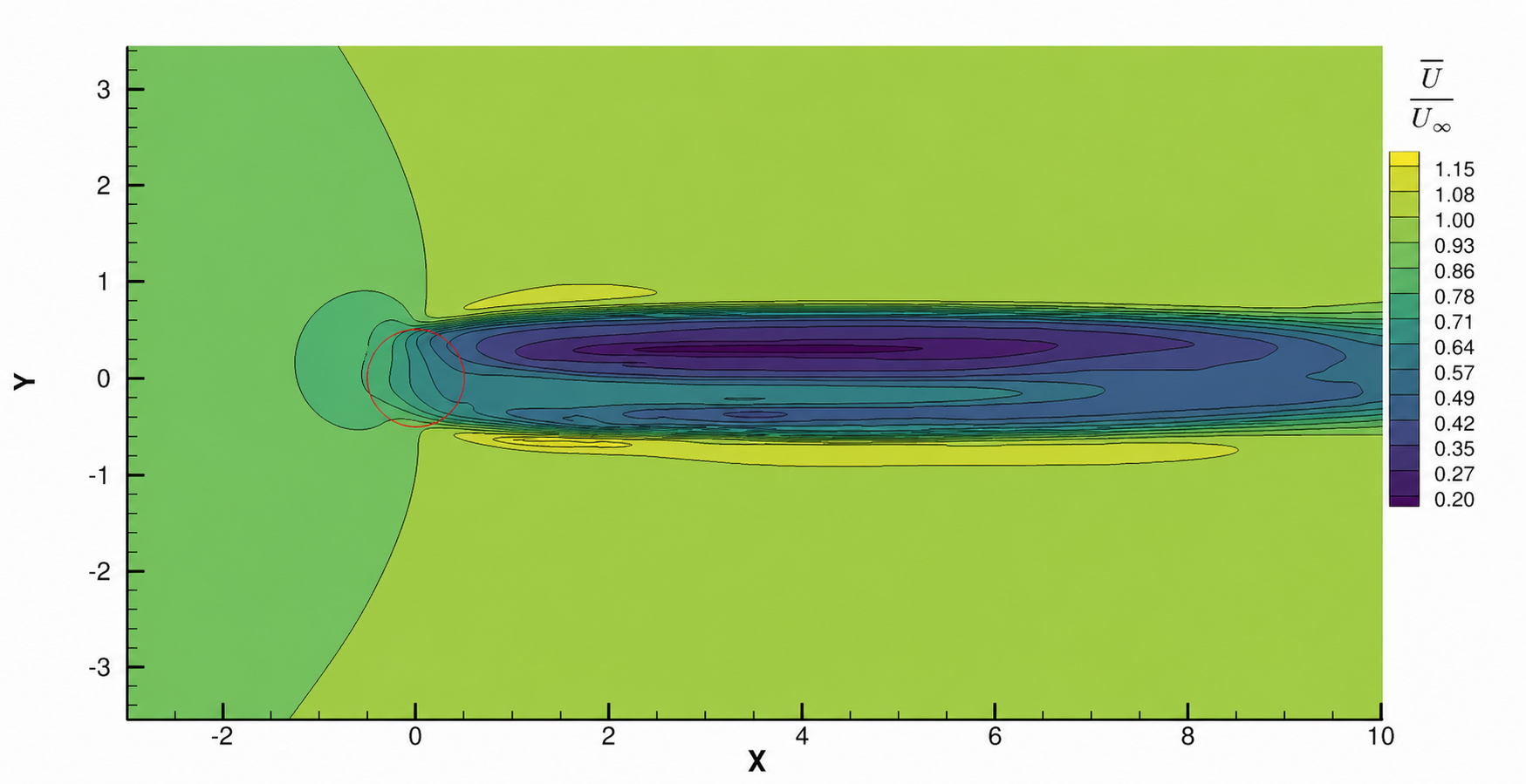}
    \caption{Mean normalized streamwise velocity}
    \label{fig:mean_u}
  \end{subfigure}\hfill
  \begin{subfigure}[b]{0.49\linewidth}
    \centering
    \includegraphics[width=\linewidth]{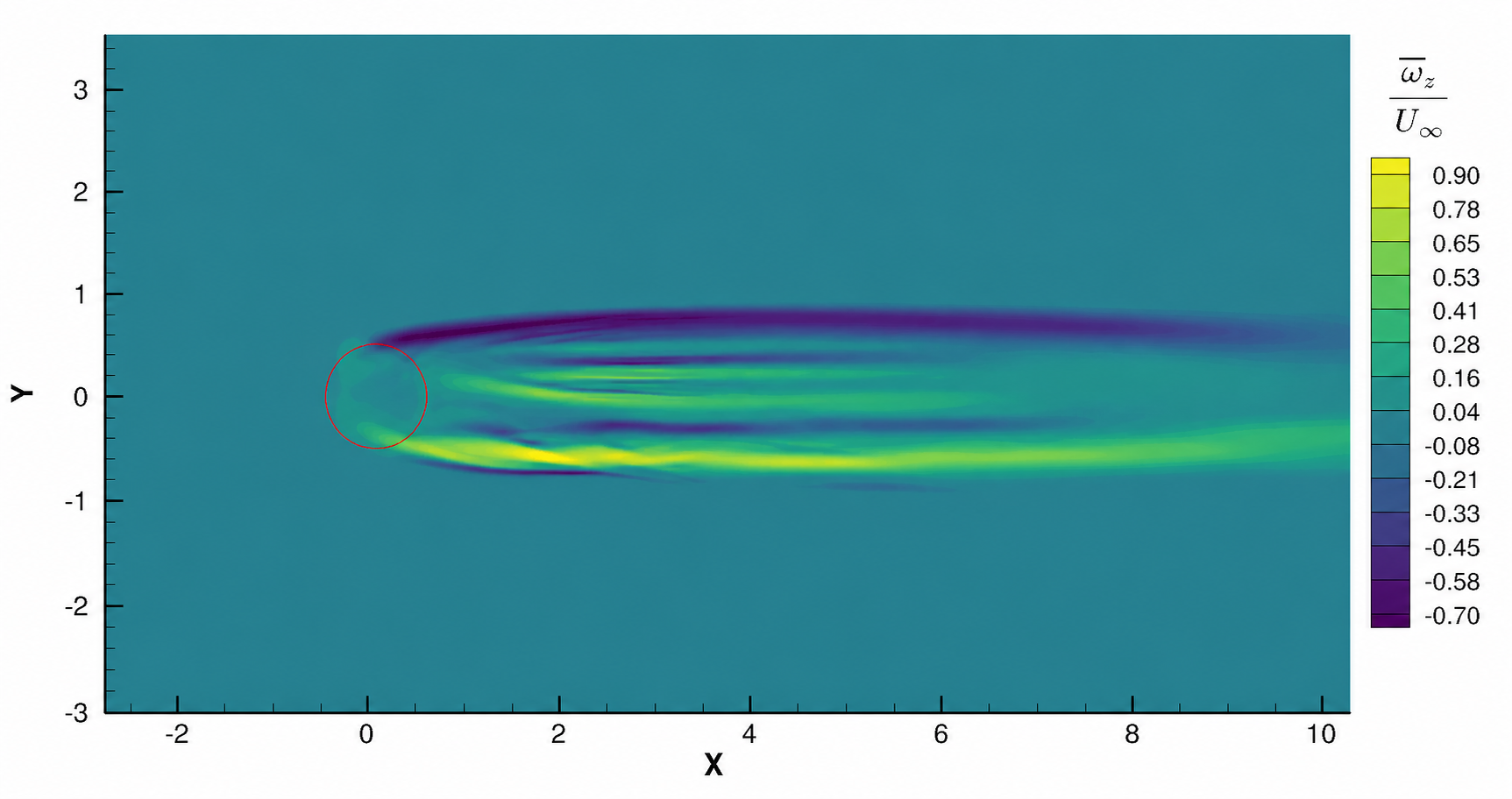}
    \caption{Mean normalized spanwise vorticity}
    \label{fig:mean_vortz}
  \end{subfigure}
  \caption{Time-averaged flowfield at $\mathrm{TSR}=2$ highlighting the persistent wake velocity deficit and the mean vorticity structure resulting from unsteady blade-wake interactions. The circular outline indicates the projected turbine swept area $D=1$.}
  \label{fig:mean_fields}
\end{figure*}

Figure~\ref{fig:inst_all}(a--b) shows the instantaneous normalized streamwise velocity and spanwise vorticity. A clear velocity deficit forms immediately downstream of the rotor, while the vorticity field reveals coherent structures shed from the blade locations and convected into the wake. The wake is highly unsteady, with strong shear layers and vortex interactions developing close to the rotor.

Figure~\ref{fig:inst_all}(c--d) shows the corresponding momentum source terms $S_x$ and $S_y$ used to represent the blade forces. The forcing is concentrated at the instantaneous blade positions and smoothly distributed due to the Gaussian kernel. The streamwise component $S_x$ is predominantly negative, indicating the extraction of momentum from the incoming flow and directly contributing to the wake deficit. The cross-stream component $S_y$ changes sign depending on the blade orientation and represents the lift-induced forcing responsible for vortex generation and wake deflection. Comparing Fig.~\ref{fig:inst_all}(a--d), it can be seen that the localized forcing regions coincide with the origin of the wake structures. The actuator-line model therefore captures the blade–flow interaction by introducing moving regions of momentum forcing, which generate both the velocity deficit and the vortex-dominated wake observed in the instantaneous flow.

The time-averaged fields are shown separately in Fig.~\ref{fig:mean_fields}. The mean streamwise velocity exhibits a sustained wake deficit with gradual downstream recovery and lateral spreading. The mean vorticity field retains residual structures along the wake shear layers, indicating that the unsteady blade-induced vortices leave a persistent imprint on the averaged flow. These mean-flow features form the basis for the wake analysis presented in the next subsection.

\subsection{Wake Profiles}
\begin{figure}[!]
\centering
\includegraphics[width=\linewidth]{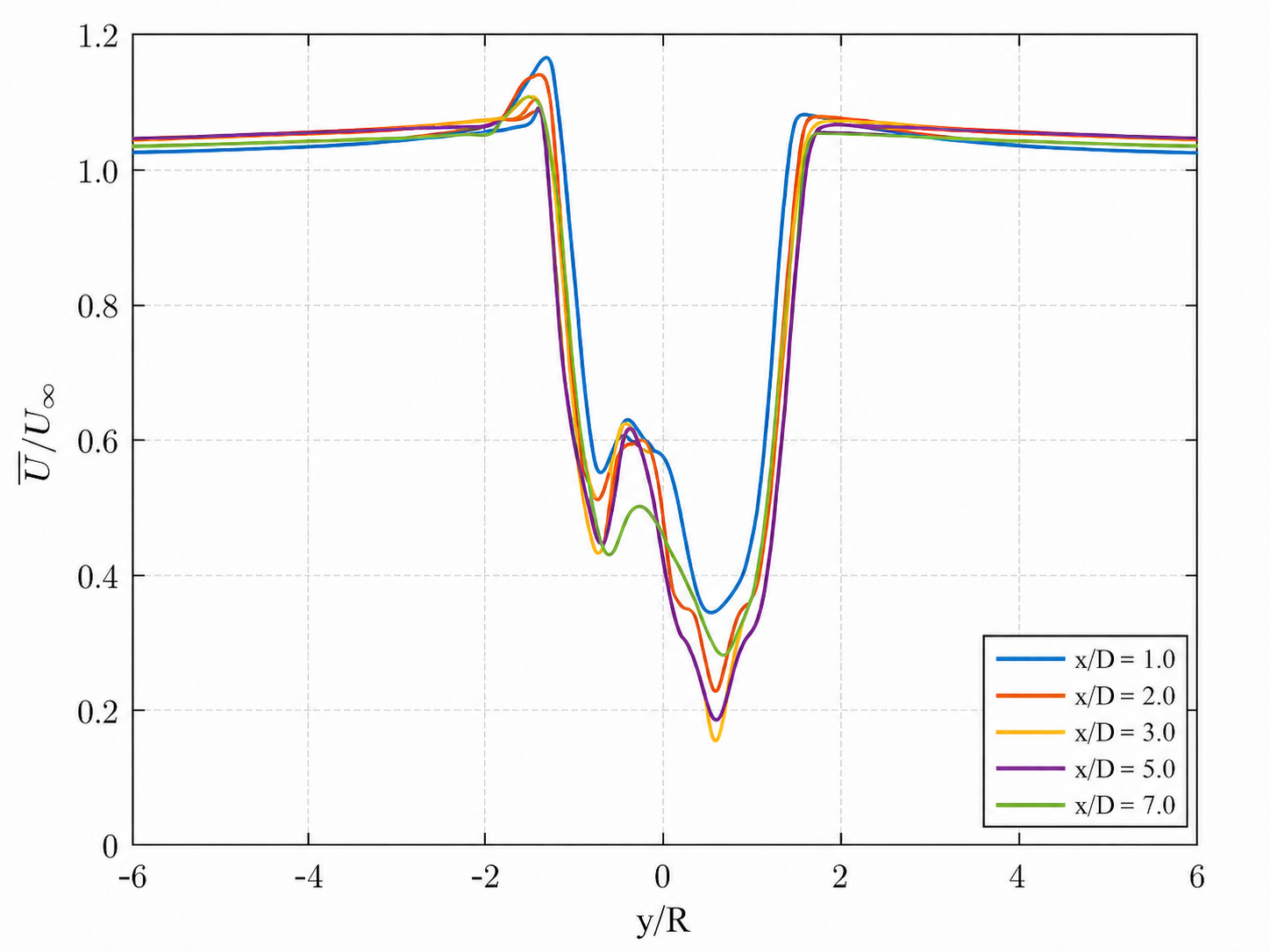}
\caption{Normalized mean streamwise velocity profiles at several downstream locations in the wake.}
\label{fig:wake_profiles}
\end{figure}

The wake structure is further examined using normalized mean streamwise velocity profiles extracted from the time-averaged flow field. Figure~\ref{fig:wake_profiles} shows $U/U_\infty$ as a function of the normalized cross-stream coordinate $y/R$ at several downstream locations $x/D$. The profiles exhibit a pronounced velocity deficit near the wake centerline and gradual recovery toward the freestream with increasing $|y/R|$, indicating the lateral spreading of the wake. Moving downstream, the wake deficit becomes progressively weaker while the wake width increases, reflecting the gradual recovery of the mean flow.

\section{Conclusions}
We investigated the feasibility of coupling a high-order FR/CPR framework with a virtual-body ALM formulation for simulating rotating VAWT on fixed cartesian meshes. Using a two-bladed turbine operating at $\mathrm{TSR}=2$, the proposed approach was evaluated under a highly unsteady and challenging operating regime. The results confirm that the combined high-order FR/CPR and ALM framework can robustly represent fully rotating turbine kinematics, unsteady blade loading trends, and the resulting wake response without the need for body-fitted meshes. Although the use of quasi-steady airfoil data limits the resolution of detailed dynamic stall effects, the framework captures the dominant flow and wake characteristics in a stable and computationally efficient manner. The primary contribution of this work is the development of a flexible high-order numerical framework for actuator-based turbine simulations. The present approach also provides a basis for future extensions to multi-turbine configurations and three-dimensional rotor–wake interaction studies.



\bibliographystyle{asmeconf}  
\bibliography{asmeconf-sample}

\appendix




\end{document}